# Portfolio Optimization – A Comparative Study

Jaydip Sen* and Subhasis Dasgupta
Department of Data Science, Praxis Business School, Kolkata, India.
*Corresponding author email: jaydip.sen@acm.org.

**Abstract**

Portfolio optimization has been an area that has attracted considerable attention from the financial research community. Designing a profitable portfolio is a challenging task involving precise forecasting of future stock returns and risks. This chapter presents a comparative study of three portfolio design approaches, the mean-variance portfolio (MVP), hierarchical risk parity (HRP)-based portfolio, and autoencoder-based portfolio. These three approaches to portfolio design are applied to the historical prices of stocks chosen from ten thematic sectors listed on the National Stock Exchange (NSE) of India. The portfolios are designed using the stock price data from January 1, 2018, to December 31, 2021, and their performances are tested on the out-of-sample data from January 1, 2022, to December 31, 2022. Extensive results are analyzed on the performance of the portfolios. It is observed that the performance of the MVP portfolio is the best on the out-of-sample data for the risk-adjusted returns. However, the autoencoder portfolios outperformed their counterparts on annual returns.

## 1. Introduction

Portfolio Optimization is the task of identifying a set of capital assets and their respective weights of allocation, which optimizes the risk-return pairs. Optimizing a portfolio is a computationally hard problem. The problem gets more complicated if one needs to optimize future return and risk values, as predicting future stock prices is equally challenging. Markowitz proposed the mean-variance optimization approach which is based on the mean and covariance matrix of returns [1]. However, the mean-variance portfolio (MVP) design poses several challenges including the difficulty in estimating future expected returns of the stocks constituting the portfolio.

The hierarchical risk parity (HRP) algorithm proposed by de Prado attempts to address the challenges of the quadratic optimization problem which are relevant to the MVP portfolio design [2]. The HRP approach to portfolio design involves clustering the stocks using an agglomerative clustering method. The weights of the stocks in a cluster are assigned in inverse proportion to the variance of the cluster. The HRP algorithm is built on the concepts of graph theory and machine learning, and unlike the MVP approach to portfolio optimization, it does not require the invertibility of the covariance matrix of the stock returns [2].

Autoencoders are symmetric networks used for unsupervised learning. The output layer of an autoencoder is of the same size as the input layer because its purpose is to reconstruct its own inputs rather than predict a dependent target value. The goal of these networks is to act as a compression filter via an encoding layer that fits the input vector into a smaller latent representation. A decoding layer reconstructs the input while minimizing the error in reconstruction. Autoencoders can be trained on the historical prices of stocks forming a portfolio, in which the salient features of the stocks are represented in a compact manner at the coding layer, while the final output layer represents the reconstructed features of the stocks.

This work presented in this chapter discusses an algorithmic approach to building optimized portfolios by selecting stocks from ten thematic sectors of the National Stock Exchange (NSE) of India. Based on the report of the NSE on December 31, 2021, ten stocks have been identified which have the highest free-float market capitalization as per their listing in the National Stock Exchange (NSE) [3]. The historical prices of these stocks are scraped from the web using their ticker names. MVP, HRP, and autoencoder-based portfolios are designed and trained using the historical prices of the stocks from January 1, 2018, to December 31, 2021. The testing of the portfolios is done on the stock price data from January 1, 2022, to December 31, 2022. Extensive analysis of the performance of the portfolios is made based on their annual returns, annual volatilities, and Sharpe ratios.

The main contribution of the current work is threefold. First, it presents three different methods of optimizing portfolios, MVP, HRP, and autoencoder-based design. These portfolio design approaches are applied to ten thematic sectors of stocks of the NSE. The results can be used as a guide to investors in the stock market for making profitable investments. Second, a method is presented for evaluating the performance of the portfolios based on their annual returns and risks and Sharpe ratios. Since the evaluation is done both on the training and the test data, the work has identified the most efficient portfolio for all ten sectors on both datasets. Hence, a robust framework for evaluating different portfolios is demonstrated. Third, the returns of the portfolios on the thematic sectors on the test data highlight the current profitability and the volatility of these sectors. This information can be useful for investors.

The chapter is organized as follows. Section 2 briefly discusses some related works in the literature. Section 3 provides a brief theoretical foundation of MVP, HRP, and autoencoder-based portfolio design. Section 4 discusses the data used and the methodology followed in the work. Section 5 presents extensive results of the performance of the portfolios and their analysis. A comparative study of the performance of the portfolios is also made. Section 6 concludes the chapter and identifies some future directions of work.

## 2. Related Work

Several approaches have been proposed by researchers for accurate prediction of stock prices and robust portfolio optimization. Time series decomposition and econometric approaches like ARIMA, Granger causality, and VAR are extensively used for stock price prediction and portfolio optimization [4-10].

The use of machine learning, deep learning, and reinforcement learning models for future stock price prediction has been the most popular approach of late [11-23]. Hybrid models are also proposed that utilize the algorithms and architectures of machine learning and deep learning and exploit the sentiments in the textual sources on the social web [24-29].

The use of metaheuristics algorithms in solving multi-objective optimization problems for portfolio management has been proposed in several works [30-32]. The use of fuzzy logic, genetic algorithms (GAs), and algorithms of swarm intelligence (SI), e.g., particle swarm optimization (PSO), are also quite common in portfolio optimization [33-35].

The performances of the mean-variance, Eigen, and HRP portfolios have been compared on different stocks from various sectors of the Indian stock market [36-42]. A pair portfolio design approach using cointegration for the Indian stock market has also been proposed in the literature [43]. The use of generalized autoregressive conditional heteroscedasticity (GARCH) in estimating the future volatility of stocks and portfolios has also been illustrated [44-45].

Finally, deep reinforcement learning approaches have been extensively used in portfolio optimization [46-58].

## 3. Theoretical Background

In this section, some background theories of portfolio design are discussed. The design approaches to MVP, HRP, and autoencoder-based portfolios and their optimization methods are briefly presented in the following.

### 3.1 Mean-Variance Portfolio Optimization

The mean-variance portfolio (MVP) design involves the following steps: (a)Computation of returns and volatilities of the stocks, (b) Determination of the covariances and correlations among all pairs of stocks in the portfolio, (c) Derivation of the expected returns and risks of several candidate portfolios, (d) Identifying the portfolio with the maximum risk-adjusted return among all candidate portfolios. In the following, we briefly describe these steps. For further details, interested readers may refer to [36].

***Computation of Stock Returns and Volatilities***: Based on the historical values of the stock prices in the portfolio training dataset, the daily return or the log return values of each stock of that sector are computed. The daily return values are the percentage changes in the successive daily stock prices, while the log return values are the logarithms of the percentage changes in the daily stock prices. Based on the daily return values, the daily volatility, and the annual volatility of every stock in a portfolio are computed. The daily volatility is the standard deviation of the daily return values. The daily volatility, on multiplication by a factor of the square root of 250, yields the value of the annual volatility. Here, there is a standard assumption of 250 working days in a year for a stock market. The annual volatility of a stock reflects the risk associated with stock from an investor's point of view. For every stock, the daily return values are also aggregated into their annual return values.

***Determination of Covariances and Correlations of Stocks***: After the volatility and return of the stocks are computed based on the historical prices of the stocks, the covariance and the correlation matrices are derived for a portfolio. These matrices depict the strength of association between all pairs of stocks in a portfolio. A good portfolio aims to minimize the risk while maximizing the return. Risk minimization of a portfolio requires identifying stocks that have low correlation among themselves so that a higher diversity can be achieved.

***Derivation of Portfolio Returns and Risks***: The expected return $E(R)$ of a portfolio containing $n$ stocks denoted as $S_1$, $S_2$,......, $S_n$, and their corresponding associated weights $w_1$, $w_2$,......,$w_n$ is given by (1):

$$E(R) = w_1 E(R_{S_1}) + w_2 E(R_{S_2}) + \cdots + w_n E(R_{S_n}) \qquad (1)$$

The variance of a portfolio is computed using the variances of the individual stocks constituting the portfolio, and the covariances between every pair of stocks in the portfolio. The variance of a portfolio, *Var(P)* is computed using (2):

$$Var(P) = \sum_{i=1}^{n} w_i \sigma_i^2 + 2 * \sum_{i,j} w_i * w_j * Cov(i,j) \quad (2)$$

In (2), $w_i$ and $\sigma_i$ represent the weight associated with stock *i* and the standard deviation of the historical prices of stock *i*. The covariance between the historical prices of stock *i* and stock *j* is denoted as $Cov(i,j)$.

Once the return and the volatility (i.e., risk) for a portfolio are computed using the stock price data of all its constituent stocks, the portfolio is optimized so that its risk-adjusted return is the maximum. This optimization is carried out in the next step.

***Portfolio with Maximum Risk-Adjusted Return***: To understand how the portfolio with the maximum-risk-adjusted return (i.e., the optimum portfolio) is identified, two concepts are important to know, (i) Sharpe ratio and (ii) efficient frontier of portfolios. In the following, these two terms are explained first.

The Sharpe ratio (SR) of a portfolio is given by (3)

$$SR = \frac{R_c - R_f}{\sigma_c} \quad (3)$$

In (3), $R_c$, $R_f$, and $\sigma_c$ denote the return of the current portfolio, the risk-free portfolio, and the standard deviation of the current portfolio, respectively. Here, the risk-free portfolio is a portfolio with a volatility value of 1%. The optimum portfolio is the one that maximizes the Sharpe Ratio for a set of stocks.

The question that remains to be addressed is how to identify the portfolio with the maximum Sharpe ratio. To answer this question, the term *efficient frontier* needs to be introduced.

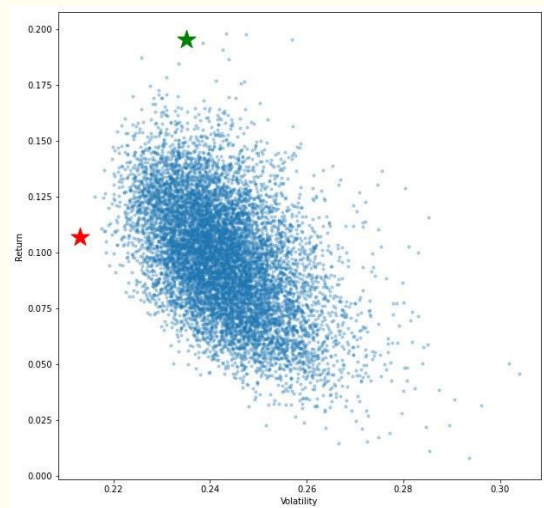

**Figure 1:** The efficient frontier illustrated with 10000 candidate portfolios. The portfolio with the minimum risk is represented by the red star, while the green star identifies the portfolio with the maximum Sharpe ratio (i.e., it is the optimum portfolio)

For a given portfolio of stocks, the *efficient frontier* is the contour with the returns plotted along the *y*-axis and the volatility (i.e., risk) on the *x*-axis. The points of an efficient frontier indicate the portfolios with the maximum return for a given value of volatility, or those with the minimum value of volatility for a given value of the return. Since, for an efficient frontier, the volatility is plotted along the *x*-axis, the minimum risk portfolio is identified by the leftmost point on the efficient frontier. Since the optimum portfolio is the one that maximizes the Sharpe ratio, this portfolio is identified by the point on the efficient frontier, that yields that maximum value of the return/risk ratio. Figure 1 depicts the efficient frontier for many candidate portfolios, wherein the portfolio with the minimum risk and the one with the maximum Sharpe ratio are identified.

### 3.2 Hierarchical Risk Parity-Based Portfolio Optimization

The execution of the hierarchical risk parity (HRP) approach to portfolio optimization involves three steps: (a) Formation of clusters, (b) Quasi-diagonalization, and (c) Recursive Bisection. These steps are briefly described below. For more details, the readers may refer to [2, 59].

***Formation of Clusters***: The tree clustering used in the HRP algorithm is an agglomerative clustering algorithm. A hierarchy class is first created in Python to design the agglomerative clustering algorithm. The hierarchy class contains a dendrogram method that receives the value returned by a method called linkage defined in the same class. The linkage method receives the dataset after pre-processing and transformation and computes the minimum distances between stocks based on their return values. There are several options for computing distance. However, the ward distance is a good choice since it minimizes the variances in the distance between two clusters in the presence of high volatility in the stock return values. In this work, the ward distance has been used as a method to compute the distance between two clusters. The linkage method performs the clustering and returns a list of the clusters formed. The computation of linkages is followed by the visualization of the clusters through a dendrogram. In the dendrogram, the leaves represent the individual stocks, while the root depicts the cluster containing all the stocks. The distance between each cluster formed is represented along the y-axis, longer arms indicate less correlated clusters and vice versa. The details of the clustering process are described in [2]. Figure 2 exhibits a typical dendrogram of the agglomerative clustering used in HRP portfolio optimization.

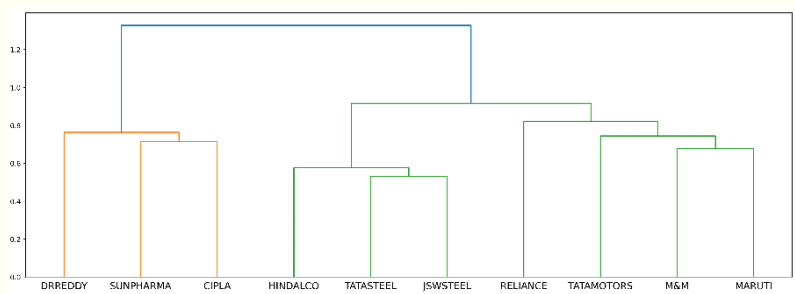

**Figure 2:** The dendrogram produced by the agglomerative clustering done by the hierarchical risk parity-based approach to portfolio optimization. The x-axis shows the ten stocks that participated in the clustering process, while the y-axis depicts the ward distance used as the metric in computing the inter-cluster distance.

***Quasi-Diagonalization***: In this step, the rows and the columns of the covariance matrix of the return values of the stocks are reorganized in such a way that the largest values lie along the diagonal. Without requiring a change in the basis of the covariance matrix, quasi-diagonalization yields a very important property of the matrix – the assets (i.e., stocks) with similar return values are placed closer to each other, while disparate assets are put at a far distance. The working principles of the algorithm are as follows. Since each row of the linkage matrix merges two branches into one, the clusters ($C_{N-1}$, 1) and ($C_{N-2}$, 2) are replaced with their constituents recursively, until there are no more clusters to merge. This recursive merging of clusters preserves the original order of the clusters [59]. The output of the algorithm is a sorted list of the original stocks (as they were before the clustering).

***Recursive Bijection***: The quasi-diagonalization step transforms the covariance matrix into a quasi-diagonal form. It is proven mathematically that the allocation of weights to the assets in an inverse ratio to their variance is an optimal allocation for a quasi-diagonal matrix [59]. This allocation may be done in two different ways. In the bottom-up approach, the variance of a contiguous subset of stocks is computed as the variance of an inverse-variance allocation of the composite cluster. In the alternative top-down approach, the allocation among two adjacent subsets of stocks is done in inverse proportion to their aggregated variances. In the current implementation, the top-down approach is followed. A Python function *computeIVP* computes the inverse-variance portfolio based on the computed variances of two clusters as its given input. The variance of a cluster is computed using another Python function called clusterVar. The output of the *clusterVar* function is used as the input to another Python function called *recBisect* which computes the final weights allocated to the individual stocks based on the recursive bisection algorithm.

### 3.3 Autoencoder-Based Portfolio Optimization

Autoencoders are symmetric networks used for unsupervised learning. The output layer is the same size as the input layer because it aims to reconstruct its own inputs rather than predict a dependent target value. The goal of these networks is to act as a compression filter via an encoding layer, $\Phi$ that fits the input vector X into a smaller latent representation (the code) *c*, and then a decoding layer, $\varphi$ tries to reconstruct it back to X' such that (4) holds good:

$$\phi: X \to c, \varphi: c \to X' \tag{4}$$

Any reconstruction error is evaluated based on the computation of a loss function. Minimization of the loss function will force the network to find the most efficient compact representation of the training data with minimum information loss. For numerical input, the loss function ($L_{MSE}$) is the *mean squared error* computed in (5):

$$L_{MSE} = \|X - X'\|^2 \tag{5}$$

As depicted in Figure 3, central layer (the code) of the network is the compressed representation of the data. We are effectively translating an *n*-dimensional array into a smaller *m*-dimensional array, where $m \leq n$. Since autoencoders can learn new latent representations, combining the previously learned ones so that each hidden level can be seen as some compressed hierarchical representation of the original data, the coding layer or any other intermediate hidden layer in the encoder part of the network can be taken as valid features describing the input vector.

Autoencoders can be trained on the historical prices of stocks forming a portfolio, in which the salient features of the stocks can be represented in a most compact manner at the coding layer, while the final output layer represents the reconstructed features of the stocks. Since for portfolio optimization, we need weights to be allocated to each stock based on the importance of the features, the extracted normalized feature values at the output layer are taken as the weights for the corresponding stocks in the portfolio [60].

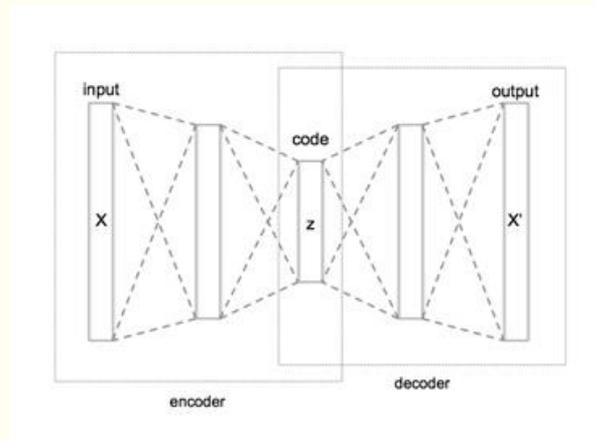

**Figure 3:** The schematic representation of an autoencoder with its input layer, encoder network, coding layer, decoding network, and output layer indicated.

## 4. Data and Methodology

**1. *Choosing the sectors*:** The stocks under ten thematic sectors of NSE, India are chosen for the portfolio design. These sectors are: (i) NIFTY commodities, (ii) NIFTY energy, (iii) NIFTY manufacturing, (iv) NIFTY services, (v) NIFTY MNC, (vi) NIFTY transportation & logistics, (vii) NIFTY infrastructure, (viii) NIFTY housing, (ix) NIFTY consumption, and (x) NIFTY 100 ESG (environmental, social and governance). The NIFTY thematic sector indices reflect the performance of stocks that belong to specific investment themes such as manufacturing, services, social, infrastructure, etc. For each sector, ten stocks are identified which have the maximum free-float market capitalization based on the NSE's report of February 29, 2022 [3].

**2. *Acquiring the data*:** The *DataReader* function defined in the *pandas* library of Python is used for scraping the historical prices of the stocks of the ten thematic sectors from the Yahoo Finance website. The stock price records for the period January 1, 2018, to December 31, 2021, are used to build the portfolios for the sectors, while the portfolios are tested on the stock records for the period January 1, 2022, to December 31, 2022. Since the current work is based on a univariate analysis, only the *close* prices of the stocks are used for computing the portfolio return and risk.

**3. *Designing the MVP portfolios*:** For designing the MVP portfolio for each sector, the daily returns of the stocks are computed using the *pct_change* function in Python. The annual return of each stock is then computed from the daily return values. The daily volatility of the stocks is computed using the *std* function in Python. The annual volatility is derived by multiplying the daily volatility by a factor of the square root of 250 assuming that there are 250 working days in a calendar year. Python functions *cov* and *corr* are used for computing the covariance and correlation matrices, respectively for each portfolio. Based on the return and the volatility of individual stocks, the annual return and risk for the portfolios are computed. Finally, For plotting the contour of the efficient frontier, the weights are assigned randomly to the ten stocks in a portfolio in a loop and iterate the loop 10,000 times in a Python program. The iteration produces 10,000 points, each point representing a portfolio.

The optimum portfolio is identified as the portfolio yielding the highest value of the Sharpe ratio on the efficient frontier.

**4. *Designing the HRP portfolios*:** For the ten sectors, the HRP portfolios are designed following the three steps, agglomerative clustering, quasi-diagonalization, and recursive bijection. The procedural and implementation details of HRP portfolios have been discussed in Section 3.2.

**5. *Designing the autoencoder portfolios*:** The fundamentals of autoencoder portfolios have been presented in Section 3.3. Here, we discuss some implementation details of the autoencoder portfolios. The model architecture of the autoencoder portfolio as produced by the *plot_model* function in the *keras* library is exhibited in Figure 4. The model has one coding layer that extracts 5 features from the 10 features supplied in the input layer. The data shape (None, 10) at the input layer represents the *close* prices of the ten stocks of the portfolio. The data shape (None, 10) at the output of the final layer corresponds to the *weights* assigned by the portfolio to the ten stocks. While the *ReLU activation* has been used at the coding layer, at the output layer, *linear activation* is used. Adam optimizer is used in model training to ensure faster convergence. The model is trained over 500 epochs using a batch size of 10. The number of layers in the autoencoder model is determined using the grid search method.

**6. *Evaluating the portfolio performance*:** Finally, the performances of the three portfolios for each of the ten sectors are evaluated on the training and test data. The metrics used for evaluation are the annual return, annual volatility, and the Sharpe ratio. For each sector, the portfolio yielding the best results on the training and test data are identified. From the point of view of the investors, the portfolio performing the best on the test data is the one that should be followed.

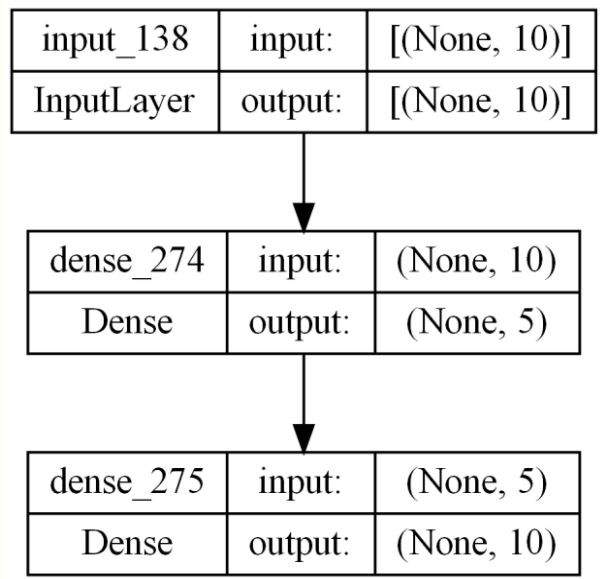

**Figure 4:** Architecture of the autoencoder model for portfolio optimization

## 5. Performance Results

This section presents the performance results of the portfolios. The portfolios are implemented using the Python language and its libraries. All experiments are carried out on a system with an Intel i7 CPU with a clock frequency in the range of 2.60 GHz

– 2.56 GHz and 16GB RAM. There are 1236 records in total, out of which 988 records are in the training dataset. The remaining 248 records are used as the test samples.

**5.1  NIFTY Commodities sector**

The ten stocks from the auto sector with the maximum free-float market capitalization and their respective contributions to the computation of the NIFTY commodities sector index according to the report published by the NSE on Dec 31, 2021, are as follows: Reliance Industries (RELIANCE): 10.13, UltraTech Cement (ULTRACEMCO): 7.52, Tata Steel (TATASTEEL): 7.52, NTPC (NTPC): 7.26, JSW Steel (JSWSTEEL): 5.64, Oil & Natural Gas Corporation (ONGC): 5.32, Grasim Industries (GRASIM): 5.31, Hindalco Industries (HINDALCO): 5.23, Coal India (COALINDIA): 4.05, and UPL (UPL): 3.32 [3]. The figures mentioned along with the names of the stocks represent the respective weights (in percent) of the stocks used in computing the sectoral index of the commodities sector. The ticker names of the stocks are mentioned within parentheses in upper case.

The weights assigned by the MVP, HRP, and the autoencoder (ENC) portfolio based on the training data (January 1, 2018 – December 31, 2021) are presented in Table 1. Figure 5 depicts the portfolio weights allocated by the portfolios in the form of pie charts. NTPC received the maximum weights from the MVP and the HRP portfolios. The ENC portfolio assigned the maximum weight to JSWSTEEL.

**Table 1: NIFTY Commodities Sector Portfolio Weights Allocation**

| Stock | Portfolio Weights | | |
|---|---|---|---|
| | MVP | HRP | ENC |
| RELIANCE | 0.2144 | 0.1464 | 0.0807 |
| ULTRACEMCO | 0.2140 | 0.1524 | 0.1079 |
| TATASTEEL | 0.0064 | 0.0428 | 0.1038 |
| NTPC | 0.3286 | 0.1803 | 0.1037 |
| JSWSTEEL | 0.0071 | 0.0459 | 0.1381 |
| ONGC | 0.0044 | 0.0614 | 0.0780 |
| GRASIM | 0.0234 | 0.1127 | 0.0997 |
| HINDALCO | 0.0025 | 0.0667 | 0.1194 |
| COALINDIA | 0.1408 | 0.0889 | 0.0590 |
| UPL | 0.0584 | 0.1025 | 0.1098 |

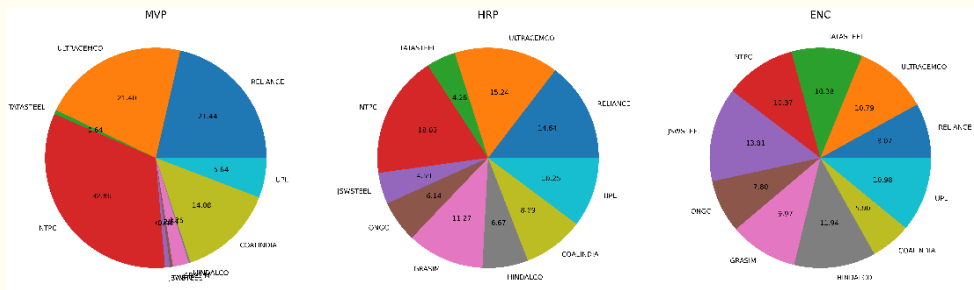

**Figure 5:** Weight allocation to the stocks of the NIFTY commodity sector by the MVP, HRP and autoencoder (ENC) portfolios

Figure 6 and Figure 7 show the cumulative daily returns of the portfolios over the training and the test periods, respectively. These plots depict the cumulative daily returns of the portfolios. The portfolio yielding a higher cumulative return is more profitable for the investors. However, the returns need to be adjusted by their associated risks. Hence, the portfolio risks and the values of their Sharpe ratios are computed so that the performances of the portfolios can be compared based on their respective Sharpe ratios.

In Table 2, the summary of the performances of the two portfolios of the au-to sector is presented for the training and the test periods. For both training and test periods, the annual returns, annual volatilities (i.e., standard deviations), and the max Sharpe ratio are tabulated in Table 2. The RL portfolio has yielded the highest Sharpe ratios for both training and test data for the auto sector.

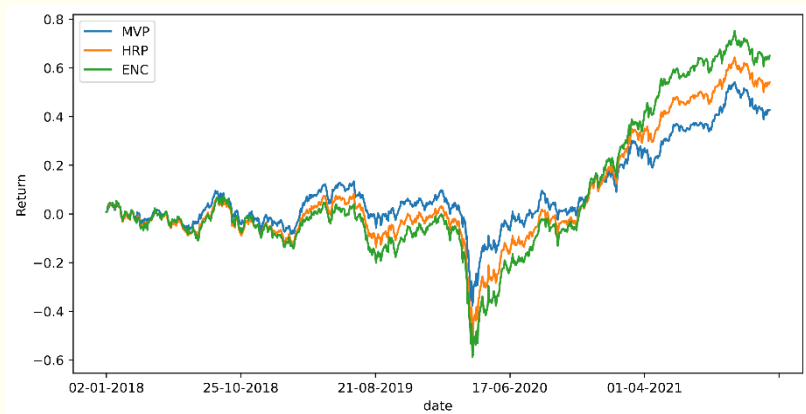

**Figure 6:** Cumulative daily returns yielded by the NIFTY commodities sector portfolios for the training period from January 1, 2018, to December 31, 2021

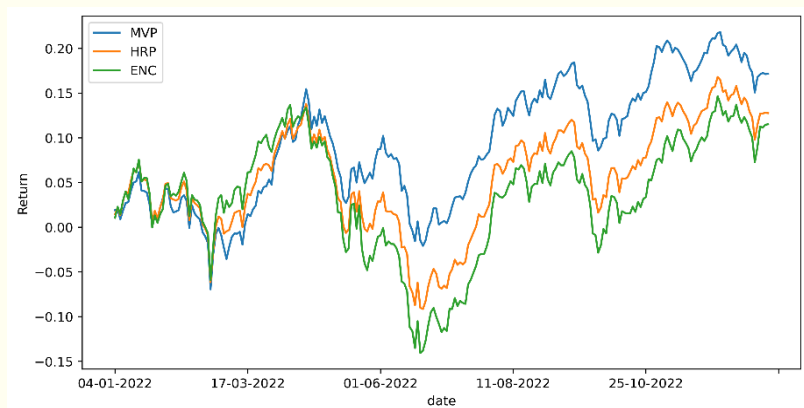

**Figure 7:** Cumulative daily returns yielded by the NIFTY commodities sector portfolios for the test period from January 1, 2022, to December 31, 2022

**Table 2: Portfolio Performance on the NIFTY Commodities Sector**

| Portfolio | Training Performance | | | Test Performance | | |
|---|---|---|---|---|---|---|
| | Annual Return | Annual Volatility | Sharpe Ratio | Annual Return | Annual Volatility | Sharpe Ratio |
| **MVP** | 10.89 | 21.87 | 0.4978 | 17.51 | 19.49 | 0.8982 |
| **HRP** | 13.82 | 23.49 | 0.5883 | 13.01 | 20.48 | 0.6354 |
| **ENC** | 16.59 | 26.01 | 0.6375 | 11.74 | 23.32 | 0.5034 |

## 5.2 NIFTY Energy sector

The ten stocks from the NIFTY Energy sector with the maximum free-float market capitalization and their respective contributions to the computation of the NIFTY commodities sector index according to the report published by the NSE on Dec 31, 2021, are as follows: Reliance Industries (RELIANCE): 35.92, NTPC (NTPC): 13.92, Power Grid Corporation of India (POWERGRID): 13.05, Oil & Natural Gas Corporation (ONGC): 10.19, Tata Power Company (TATAPOWER): 5.89, Bharat Petroleum Corporation (BPCL): 5.32, Indian Oil Corporation (IOC): 4.98, GAIL India (GAIL): 4.75, Adani Transmission (ADANITRANS): 3.08, and Adani Green Energy (ADANIGREEN): 2.90 [3]. The figures mentioned along with the names of the stocks represent the respective weights (in percent) of the stocks used in computing the sectoral index of the commodities sector. The ticker names of the stocks are mentioned within parentheses in upper case.

The weights assigned by the MVP, HRP, and the autoencoder (ENC) portfolio based on the training data (January 1, 2018 – December 31, 2021) are presented in Table 3. Figure 8 depicts the portfolio weights allocated by the portfolios in the form of pie charts. POWERGRID received the maximum weights from all three portfolios, MVP, HRP, and ENC.

**Table 3: NIFTY Energy Sector Portfolio Weights Allocation**

| Stock | Portfolio Weights | | |
|---|---|---|---|
| | **MVP** | **HRP** | **ENC** |
| RELIANCE | 0.2090 | 0.1341 | 0.1221 |
| NTPC | 0.2069 | 0.1683 | 0.0903 |
| POWERGRID | 0.3170 | 0.2184 | 0.1324 |
| ONGC | 0.0065 | 0.0508 | 0.0863 |
| TATAPOWER | 0.0170 | 0.0827 | 0.0902 |
| BPCL | 0.0106 | 0.0462 | 0.1159 |
| IOC | 0.0643 | 0.1176 | 0.0983 |
| GAIL | 0.0571 | 0.0890 | 0.0624 |
| ADANITRANS | 0.0471 | 0.0483 | 0.0624 |
| ADANIGREEN | 0.0471 | 0.0483 | 0.0843 |

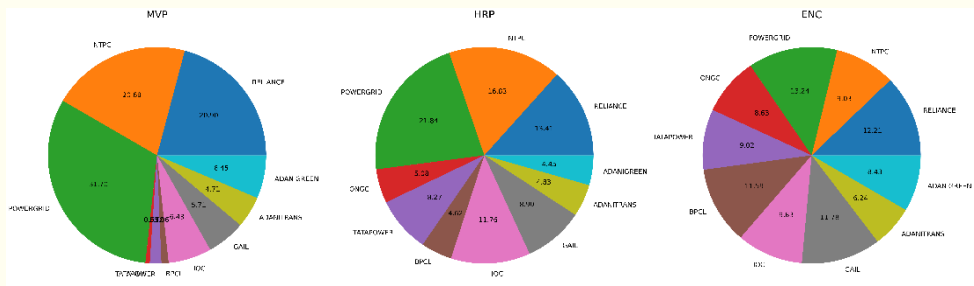

**Figure 8:** Weight allocation to the stocks of the NIFTY energy sector by the MVP, HRP and autoencoder (ENC) portfolios

Figure 9 and Figure 10 show the cumulative returns of the portfolios over the training and the test periods, respectively. In Table 4, the summary of the performances of the three portfolios of the NIFTY energy sector is presented for the training and the test periods, in which the annual returns, annual volatilities (i.e., standard deviations), and the max Sharpe ratio are tabulated.

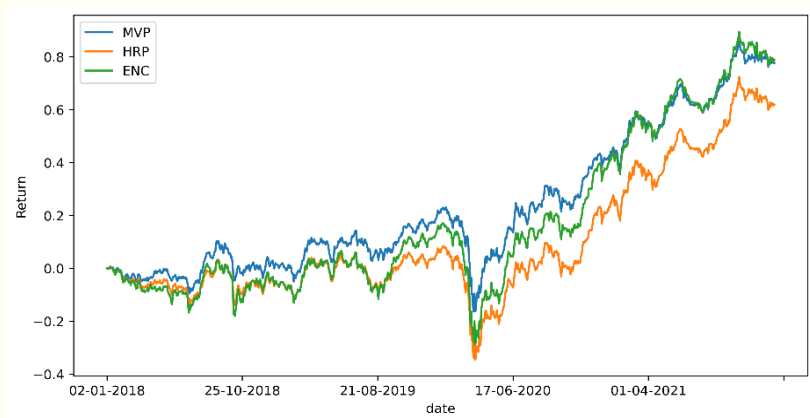

**Figure 9:** Cumulative daily returns yielded by the NIFTY energy sector portfolios for the training period from January 1, 2018, to December 31, 2021

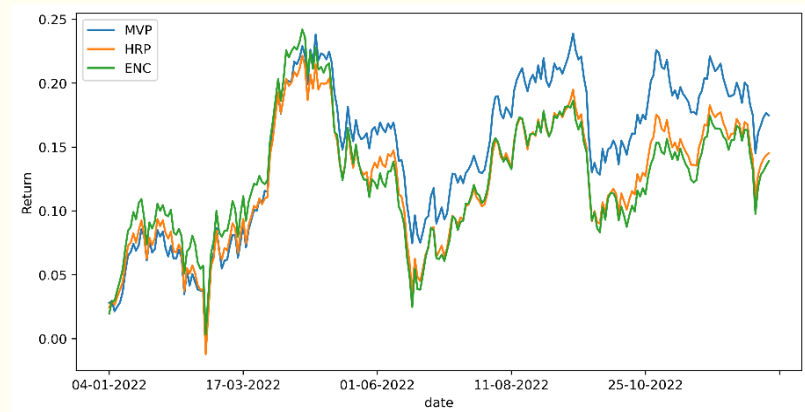

**Figure 10:** Cumulative daily returns yielded by the NIFTY energy sector portfolios for the test period from January 1, 2022, to December 31, 2022

**Table 4: Portfolio Performance in the NIFTY Energy Sector**

|  | Training Performance | | | Test Performance | | |
| --- | --- | --- | --- | --- | --- | --- |
| **Portfolio** | Annual Return | Annual Volatility | Sharpe Ratio | Annual Return | Annual Volatility | Sharpe Ratio |
| **MVP** | 19.83% | 20.98 | 0.9451 | 17.83% | 19.62 | 0.9086 |
| **HRP** | 15.81% | 21.92 | 0.7212 | 14.82% | 19.54 | 0.7583 |
| **ENC** | 20.15% | 23.31 | 0.8644 | 14.17% | 20.15 | 0.7033 |

## 5.3 NIFTY Manufacturing sector

The ten stocks from the NIFTY manufacturing sector with the maximum free-float market capitalization and their respective contributions to the computation of the overall sectoral index according to the report published by the NSE on Dec 31, 2021, are as follows: Sun Pharmaceuticals Industries (SUNPHARMA): 4.96, Reliance Industries (RELIANCE): 4.73, Mahindra & Mahindra (M&M): 4.59, Tata Steel (TATASTEEL): 4.55, Maruti Suzuki (MARUTI): 4.33, JSW Steel (JSWSTEEL): 3.41, Hindalco Industries (HINDALCO): 3.16, Tata Motors (TATAMOTORS): 2.85, Dr. Reddy's Laboratories (DRREDDY): 2.85, and Cipla (CIPLA): 2.66 [3]. The figures mentioned along with the names of the stocks represent the respective weights (in percent) of the stocks used in computing the sectoral index of the manufacturing sector. The ticker names of the stocks are mentioned within parentheses in upper case.

The weights assigned by the MVP, HRP, and autoencoder (ENC) portfolio based on the training data (January 1, 2018 – December 31, 2021) are presented in Table 5. Figure 11 depicts the portfolio weights allocated by the portfolios in the form of pie charts. DRREDDY, CIPLA, and SUNPHARMA received the highest allocation by the MVP, HRP, and ENC portfolios.

**Table 5: NIFTY Manufacturing Sector Portfolio Weights Allocation**

| Stock | Portfolio Weights | | |
|---|---|---|---|
| | MVP | HRP | ENC |
| SUNPHARMA | 0.0600 | 0.1257 | 0.1422 |
| RELIANCE | 0.1790 | 0.1077 | 0.0885 |
| M&M | 0.0738 | 0.0741 | 0.1007 |
| TATASTEEL | 0.0182 | 0.0690 | 0.0981 |
| MARUTI | 0.1437 | 0.0838 | 0.0651 |
| JSWSTEEL | 0.0130 | 0.0742 | 0.1299 |
| HINDALCO | 0.0071 | 0.0627 | 0.1044 |
| TATAMOTORS | 0.0069 | 0.0525 | 0.1167 |
| DRREDDY | 0.2629 | 0.1661 | 0.0813 |
| CIPLA | 0.2354 | 0.1840 | 0.0732 |

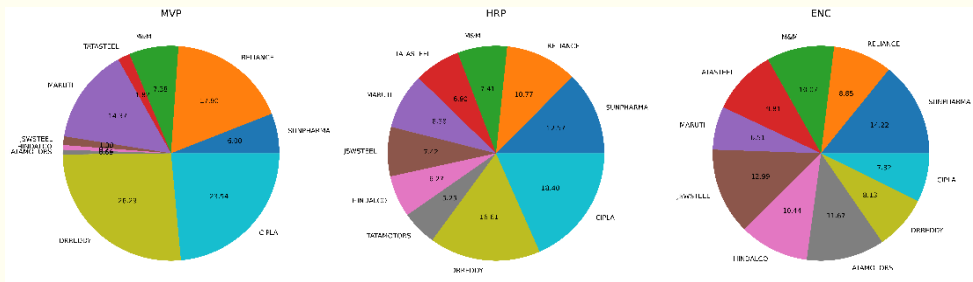

**Figure 11:** Weight allocation to the stocks of the NIFTY manufacturing sector by the MVP, HRP, and autoencoder (ENC) portfolios

Figure 12 and Figure 13 show the cumulative returns of the portfolios over the training and the test periods, respectively. In Table 6, the summary of the performances of the three portfolios of the NIFTY manufacturing sector is presented for the training and the test periods, in which the annual returns, annual volatilities (i.e., standard deviations), and the max Sharpe ratio are tabulated.

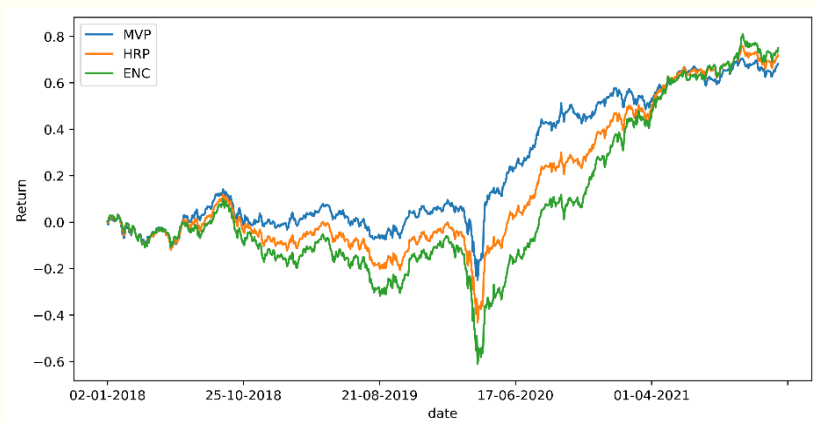

**Figure 12:** Cumulative daily returns yielded by the NIFT manufacturing sector portfolios for the training period from January 1, 2018, to December 31, 2021

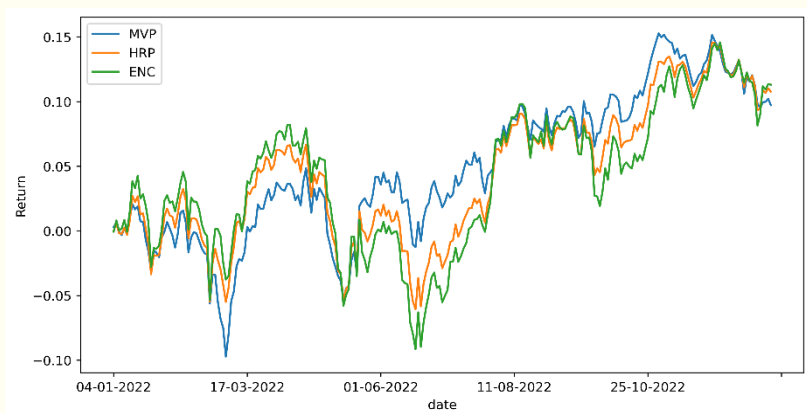

**Figure 13:** Cumulative daily returns yielded by the NIFTY manufacturing sector portfolios for the test period from January 1, 2022, to December 31, 2022

**Table 6: Portfolio Performance in the NIFTY Manufacturing Sector**

| Portfolio | Training Performance | | | Test Performance | | |
|---|---|---|---|---|---|---|
| | Annual Return | Annual Volatility | Sharpe Ratio | Annual Return | Annual Volatility | Sharpe Ratio |
| **MVP** | 17.40% | 20.76 | 0.8382 | 9.93% | 16.27 | 0.6102 |
| **HRP** | 18.40% | 22.28 | 0.8259 | 10.99% | 17.61 | 0.6242 |
| **ENC** | 19.12% | 25.58 | 0.7474 | 11.54% | 20.79 | 0.5549 |

### 5.4 NIFTY Services sector

The ten stocks from the NIFTY services sector with the maximum free-float market capitalization and their respective contributions to the computation of the overall sectoral index according to the report published by the NSE on Dec 31, 2021, are as follows: HDFC Bank (HDFCBANK): 15.11, ICICI Bank (ICICIBANK): 12.78, Infosys (INFY): 4.59, Housing Development Finance Corporation (HDFC): 10.09, Tata Consultancy Services (TCS): 7.28, Kotak Mahindra Bank (KOTAKBANK): 5.37, Axis Bank (AXISBANK): 4.90, State Bank of India (SBIN): 4.30, Bharti Airtel

(BHARTIARTL): 3.98, and Bajaj Finance (BAJFINANCE): 3.49 [3]. The figures mentioned along with the names of the stocks represent the respective weights (in percent) of the stocks used in computing the sectoral index of the NIFTY services sector.

The weights assigned by the MVP, HRP, and the autoencoder (ENC) portfolio based on the training data (January 1, 2018 – December 31, 2021) are presented in Table 7. Figure 14 depicts the portfolio weights allocated by the portfolios in the form of pie charts. TCS received the maximum weights from the MVP and HRP portfolios. The ENC portfolio assigned the highest weight to KOTAKBANK.

**Table 7: NIFTY Services Sector Portfolio Weights Allocation**

| Stock | Portfolio Weights | | |
|---|---|---|---|
| | MVP | HRP | ENC |
| HDFCBANK | 0.2593 | 0.0873 | 0.0788 |
| ICICIBANK | 0.0020 | 0.0727 | 0.0924 |
| INFY | 0.1445 | 0.1622 | 0.1194 |
| HDFC | 0.0030 | 0.0605 | 0.1194 |
| TCS | 0.3253 | 0.1933 | 0.0963 |
| KOTAKBANK | 0.0832 | 0.1096 | 0.1598 |
| AXISBANK | 0.0013 | 0.0627 | 0.0901 |
| SBIN | 0.0349 | 0.0725 | 0.0901 |
| BHARTIARTL | 0.1453 | 0.1225 | 0.0499 |
| BAJFINANCE | 0.0011 | 0.0566 | 0.0974 |

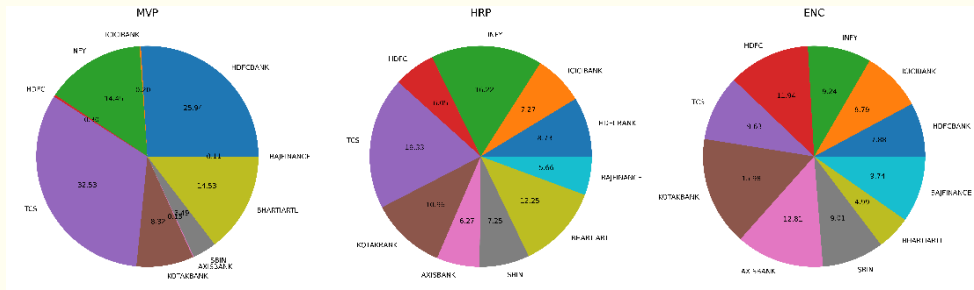

**Figure 14:** Weight allocation to the stocks of the NIFTY services sector by the MVP, HRP and autoencoder (ENC) portfolios

Figure 15 and Figure 16 show the cumulative returns of the portfolios over the training and the test periods, respectively. In Table 8, the summary of the performances of the three portfolios of the NIFTY services sector is presented for the training and the test periods, in which the annual returns, annual volatilities (i.e., standard deviations), and the max Sharpe ratio are tabulated.

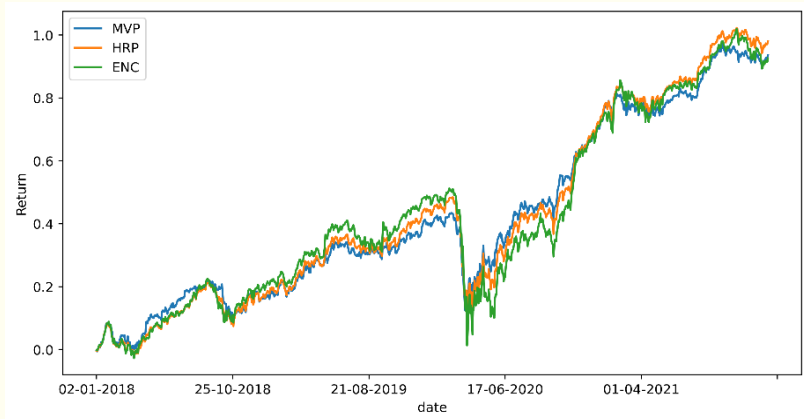

**Figure 15:** Cumulative daily returns yielded by the NIFTY services sector portfolios for the training period from January 1, 2018, to December 31, 2021

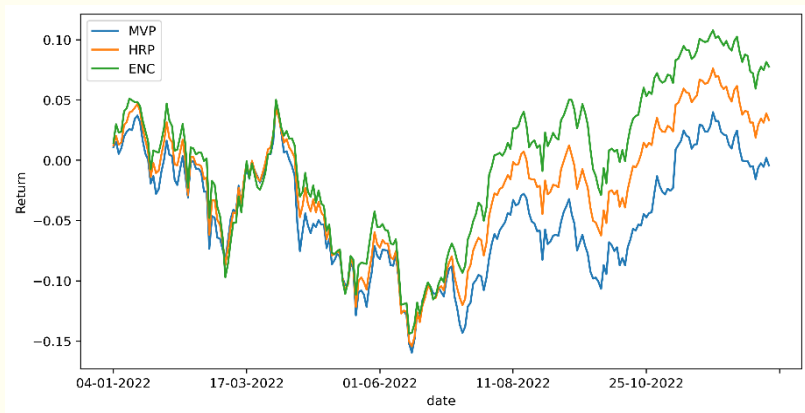

**Figure 16:** Cumulative daily returns yielded by the NIFTY services sector portfolios for the test period from January 1, 2022, to December 31, 2022

**Table 8: Portfolio Performance in the NIFTY Services Sector**

| Portfolio | Training Performance | | | Test Performance | | |
|---|---|---|---|---|---|---|
| | Annual Return | Annual Volatility | Sharpe Ratio | Annual Return | Annual Volatility | Sharpe Ratio |
| **MVP** | 23.89% | 19.81 | 1.2059 | -0.45% | 18.68 | -0.0240 |
| **HRP** | 25.03% | 21.42 | 1.1684 | 3.40% | 18.53 | 0.1834 |
| **ENC** | 23.65% | 24.31 | 0.9727 | 7.92% | 19.57 | 0.4047 |

### 5.5 NIFTY MNC sector

The ten stocks from the NIFTY MNC sector with the maximum free-float market capitalization and their respective contributions to the computation of the overall sectoral index according to the report published by the NSE on Dec 31, 2021, are as follows: Maruti Suzuki (MARUTI): 10.82, Hindustan Unilever (HINDUNILVR): 9.88, Nestle India (NESTLEIND): 9.83, Britannia Industries (BRITANNIA): 9.12, Vedanta (VEDL): 5.18, Siemens (SIEMNS): 5.00, Ambuja Cements (AMBUJACEM): 4.35, United Spirits (MCDOWELL-N): 3.82, Cummins India (CUMMINSIND): 3.69, and Ashok

Leyland (ASHOKLEY): 3.62 [3]. The figures mentioned along with the names of the stocks represent the respective weights (in percent) of the stocks used in computing the sectoral index of the NIFTY MNC sector.

The weights assigned by the MVP, HRP, and the autoencoder (ENC) portfolio based on the training data (January 1, 2018 – December 31, 2021) are presented in Table 9. Figure 17 depicts the portfolio weights allocated by the portfolios in the form of pie charts. HINDUNILVR, NESTLEIND, and VEDL received the maximum weights from the MVP, HRP, and ENC portfolios, respectively.

**Table 9: NIFTY MNC Sector Portfolio Weights Allocation**

| Stock | Portfolio Weights | | |
|---|---|---|---|
| | MVP | HRP | ENC |
| MARUTI | 0.0192 | 0.0836 | 0.0726 |
| HINDUNILVR | 0.2535 | 0.1395 | 0.0680 |
| NESTLEIND | 0.2054 | 0.1854 | 0.0715 |
| BRITANNIA | 0.1174 | 0.1219 | 0.0723 |
| VEDL | 0.0122 | 0.0605 | 0.1879 |
| SIEMENS | 0.0874 | 0.0678 | 0.1375 |
| AMBUJACEM | 0.0688 | 0.0629 | 0.0855 |
| MCDOWELL-N | 0.0769 | 0.1158 | 0.0879 |
| CUMMINSIND | 0.1541 | 0.1114 | 0.1063 |
| ASHOKLEY | 0.0050 | 0.0512 | 0.1105 |

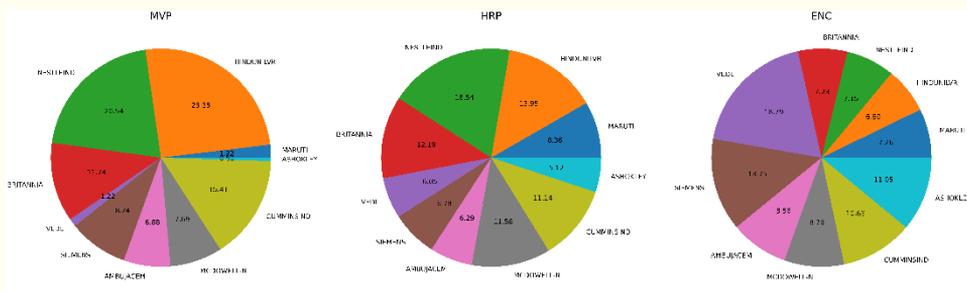

**Figure 17:** Weight allocation to the stocks of the NIFTY MNC sector by the MVP, HRP and autoencoder (ENC) portfolios

Figure 18 and Figure 19 show the cumulative returns of the portfolios over the training and the test periods, respectively. In Table 10, the summary of the performances of the three portfolios of the NIFTY MNC sector is presented for the training and the test periods, in which the annual returns, annual volatilities (i.e., standard deviations), and the max Sharpe ratio are tabulated.

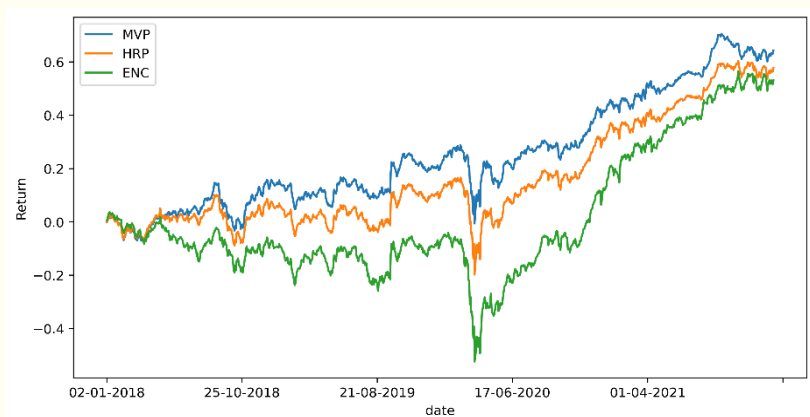

**Figure 18:** Cumulative daily returns yielded by the NIFTY MNC sector portfolios for the training period from January 1, 2018, to December 31, 2021

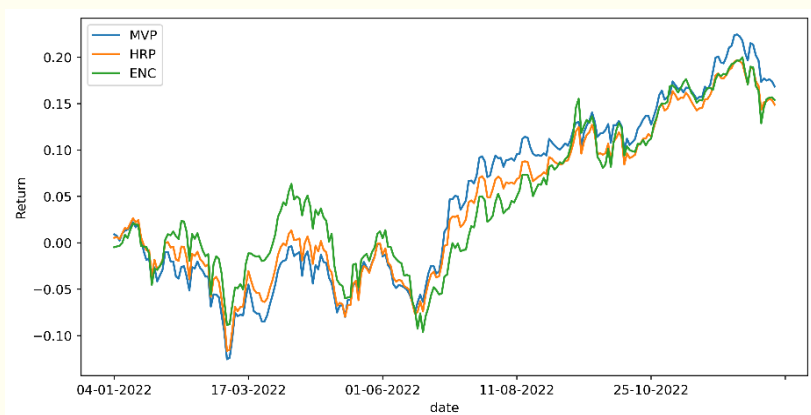

**Figure 19:** Cumulative daily returns yielded by the NIFTY MNC sector portfolios for the test period from January 1, 2022, to December 31, 2022

**Table 10: Portfolio Performance on the NIFTY MNC Sector**

| Portfolio | Training Performance | | | Test Performance | | |
|---|---|---|---|---|---|---|
| | Annual Return | Annual Volatility | Sharpe Ratio | Annual Return | Annual Volatility | Sharpe Ratio |
| **MVP** | 16.42% | 18.59 | 0.8833 | 17.18% | 17.24 | 0.9963 |
| **HRP** | 14.73% | 19.49 | 0.7557 | 15.18% | 17.33 | 0.8754 |
| **ENC** | 13.55% | 22.88 | 0.5922 | 15.70% | 20.04 | 0.7835 |

### 5.6 NIFTY Transportation and Logistics sector

The ten stocks from the NIFTY transportation and logistics sector with the maximum free-float market capitalization and their respective contributions to the computation of the overall sectoral index according to the report published by the NSE on Dec 31, 2021, are as follows: Mahindra & Mahindra (M&M): 14.73, Maruti Suzuki (MARUTI): 13.89, Tata Motors (TATAMOTORS): 9.14, Adani Ports and Special Economic Zone (ADANI PORTS): 5.43, Eicher Motors (EICHERMOT): 5.25, Bajaj Auto (BAJAJ-AUTO): 5.02, Hero MotoCorp (HEROMOTOCO): 3.81, Tube Investments of

India (TIINDIA): 3.50, TVS Motor Company (TVSMOTOR): 3.05, and Ashok Leyland (ASHOKLEY): 2.54 [3]. The figures mentioned along with the names of the stocks represent the respective weights (in percent) of the stocks used in computing the sectoral index of the NIFTY transportation and logistics sector.

The weights assigned by the MVP, HRP, and the autoencoder (ENC) portfolio based on the training data (January 1, 2018 – December 31, 2021) are presented in Table 11. Figure 20 depicts the portfolio weights allocated by the portfolios in the form of pie charts. BAJAJ-AUTO received the maximum weights from all three portfolios, MVP, HRP, and ENC.

**Table 11: NIFTY Transport & Logistics Sector Portfolio Weights Allocation**

| Stock | Portfolio Weights | | |
|---|---|---|---|
| | MVP | HRP | ENC |
| M&M | 0.0742 | 0.1207 | 0.0988 |
| MARUTI | 0.0425 | 0.1139 | 0.1028 |
| TATAMOTORS | 0.0025 | 0.0315 | 0.1081 |
| ADANIPORTS | 0.1398 | 0.1340 | 0.1020 |
| EICHERMOT | 0.0839 | 0.0800 | 0.1071 |
| BAJAJ-AUTO | 0.3344 | 0.1600 | 0.1235 |
| HEROMOTOCO | 0.0847 | 0.1237 | 0.1136 |
| TIINDIA | 0.2034 | 0.1215 | 0.0377 |
| TVSMOTOR | 0.0305 | 0.0777 | 0.1060 |
| ASHOKLEY | 0.0041 | 0.0370 | 0.1003 |

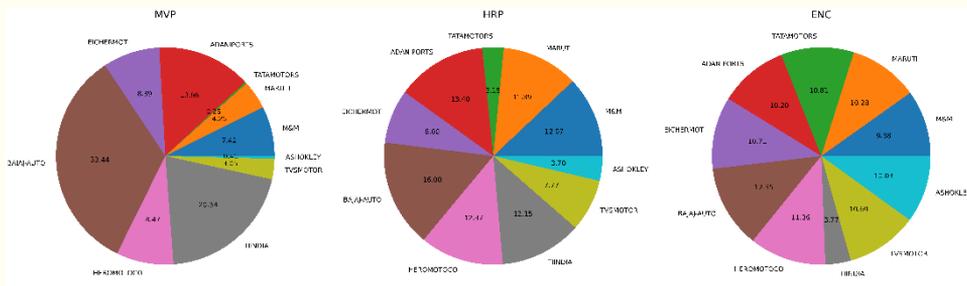

**Figure 20:** Weight allocation to the stocks of the NIFTY transportation and logistics sector by the MVP, HRP, and autoencoder (ENC) portfolios

Figure 21 and Figure 22 show the cumulative returns of the portfolios over the training and the test periods, respectively. In Table 12, the summary of the performances of the three portfolios of the NIFTY transportation and logistics sector is presented for the training and the test periods, in which the annual returns, annual volatilities (i.e., standard deviations), and the max Sharpe ratio are tabulated.

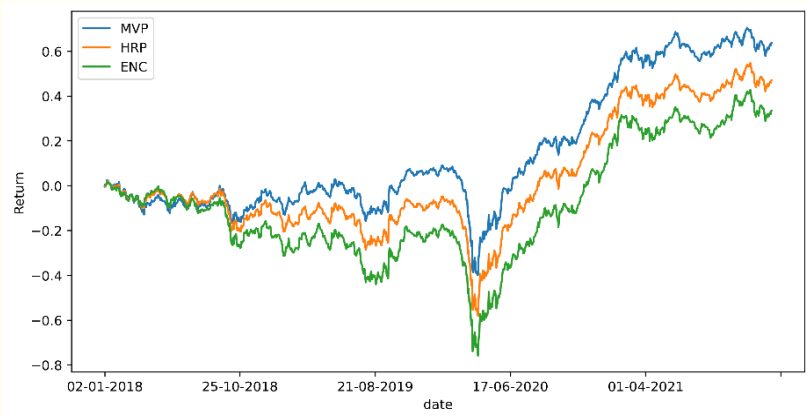

**Figure 21:** Cumulative daily returns yielded by the NIFTY transportation & logistics sector portfolios for the training period from January 1, 2018, to December 31, 2021

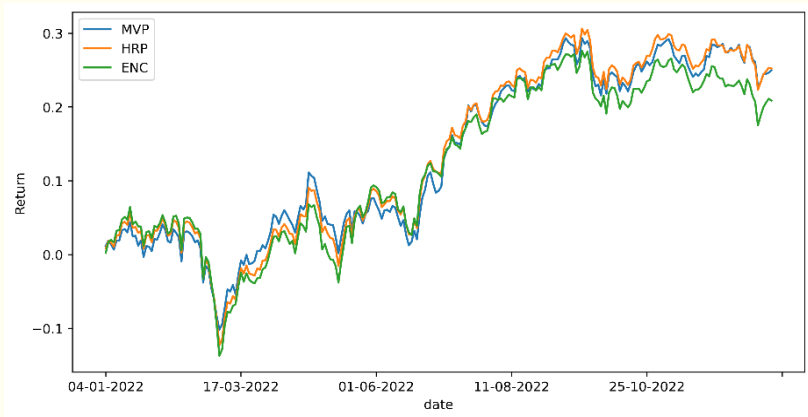

**Figure 22:** Cumulative daily returns yielded by the NIFTY transportation & logistics sector portfolios for the test period from January 1, 2022, to December 31, 2022

**Table 12: Portfolio Performance on the NIFTY Transport & Logistics Sector**

| Portfolio | Training Performance | | | Test Performance | | |
|---|---|---|---|---|---|---|
| | Annual Return | Annual Volatility | Sharpe Ratio | Annual Return | Annual Volatility | Sharpe Ratio |
| **MVP** | 16.26% | 22.75 | 0.7150 | 25.49% | 21.10 | 1.2081 |
| **HRP** | 12.02% | 23.79 | 0.5053 | 25.73% | 21.51 | 1.1962 |
| **ENC** | 8.55% | 26.03 | 0.3286 | 21.30% | 22.62 | 0.9413 |

### 5.7 NIFTY Infrastructure sector

The ten stocks from the NIFTY infrastructure sector with the maximum free-float market capitalization and their respective contributions to the computation of the overall sectoral index according to the report published by the NSE on Dec 31, 2021, are as follows: Reliance Industries (RELIANCE): 19.25, Larsen & Toubro (LT): 15.47, Bharti Airtel (BHARTIARTL): 11.28, UltraTech Cement (ULTRACEMCO): 11.28, NTPC (NTPC): 4.92, Power Grid Corporation of India (POWERGRID): 4.61, Oil & Natural Corporation (ONGC): 3.60, Grasim Industries (GRASIM): 3.59, Apollo Hospitals

Enterprise (APOLLOHOSP): 2.73, and Adani Ports and Special Economic Zone (ADANIPORTS): 2.72 [3]. The figures mentioned along with the names of the stocks represent the respective weights (in percent) of the stocks used in computing the sectoral index of the NIFTY infrastructure sector.

The weights assigned by the MVP, HRP, and the autoencoder (ENC) portfolio based on the training data (January 1, 2018 – December 31, 2021) are presented in Table 13. Figure 23 depicts the portfolio weights allocated by the portfolios in the form of pie charts. POWERGRID received the maximum weights from all three portfolios, MVP, HRP, and ENC.

**Table 13: NIFTY Infrastructure Sector Portfolio Weights Allocation**

| Stock | Portfolio Weights | | |
|---|---|---|---|
| | MVP | HRP | ENC |
| RELIANCE | 0.1152 | 0.1282 | 0.1136 |
| LT | 0.0876 | 0.1062 | 0.1274 |
| BHARTIARTL | 0.0755 | 0.1137 | 0.1088 |
| ULTRACEMCO | 0.0591 | 0.0607 | 0.0953 |
| NTPC | 0.1771 | 0.1258 | 0.1169 |
| POWERGRID | 0.3160 | 0.1480 | 0.1284 |
| ONGC | 0.0046 | 0.0434 | 0.0837 |
| GRASIM | 0.0046 | 0.0434 | 0.0837 |
| APOLLOHOSP | 0.1164 | 0.1092 | 0.0467 |
| ADANIPORTS | 0.0405 | 0.0898 | 0.0940 |

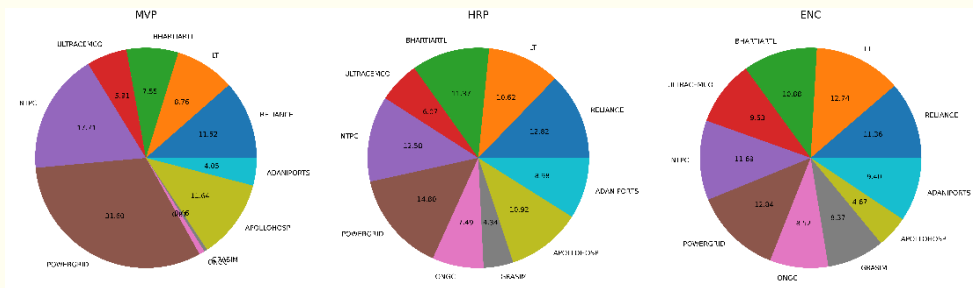

**Figure 23:** Weight allocation to the stocks of the NIFTY infrastructure sector by the MVP, HRP and autoencoder (ENC) portfolios

Figure 24 and Figure 25 show the cumulative returns of the portfolios over the training and the test periods, respectively. In Table 14, the summary of the performances of the three portfolios of the NIFTY infrastructure sector is presented for the training and the test periods, in which the annual returns, annual volatilities (i.e., standard deviations), and the max Sharpe ratio are tabulated.

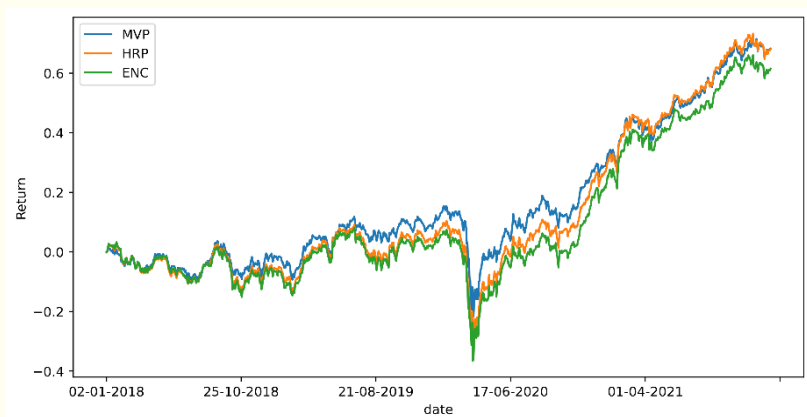

**Figure 24:** Cumulative daily returns yielded by the NIFTY infrastructure sector portfolios for the training period from January 1, 2018, to December 31, 2021

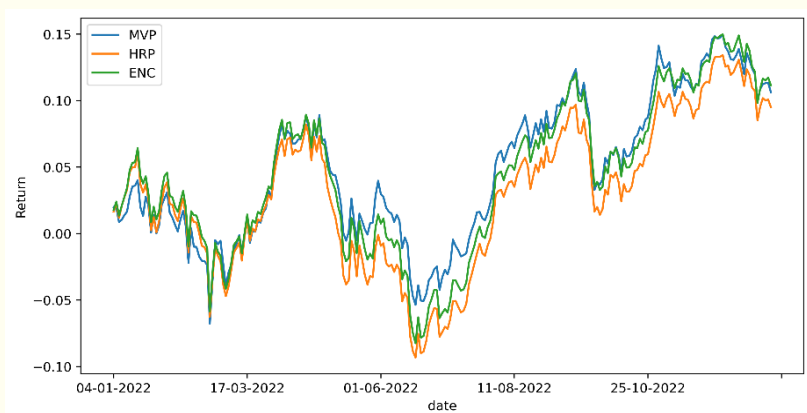

**Figure 25:** Cumulative daily returns yielded by the NIFTY infrastructure sector portfolios for the test period from January 1, 2022, to December 31, 2022

**Table 14: Portfolio Performance on the NIFTY Infrastructure Sector**

| Portfolio | Training Performance | | | Test Performance | | |
|---|---|---|---|---|---|---|
| | Annual Return | Annual Volatility | Sharpe Ratio | Annual Return | Annual Volatility | Sharpe Ratio |
| **MVP** | 17.41% | 19.78 | 0.8801 | 10.85% | 17.51 | 0.6200 |
| **HRP** | 17.42% | 20.99 | 0.8300 | 9.70% | 17.57 | 0.5523 |
| **ENC** | 15.70% | 21.11 | 0.7438 | 11.38% | 17.83 | 0.6384 |

## 5.8 NIFTY Housing sector

The ten stocks from the NIFTY housing sector with the maximum free-float market capitalization and their respective contributions to the computation of the overall sectoral index according to the report published by the NSE on Dec 31, 2021, are as follows: Larsen & Toubro (LT): 11.25, Asian Paints (ASIANPAINT): 8.59, HDFC Bank (HDFCBANK): 6.99, ICICI Bank (ICICIBANK): 5.92, UltraTech Cement (ULTRACEMCO): 5.65, Tata Steel (TATASTEEL): 5.65, NTPC (NTPC): 5.46, Housing Development Finance Corporation (HDFC): 4.67, JSW Steel (JSWSTEEL): 4.24, and

Grasim Industries (GRASIM): 3.99 [3]. The figures mentioned along with the names of the stocks represent the respective weights (in percent) of the stocks used in computing the sectoral index of the NIFTY infrastructure sector.

The weights assigned by the MVP, HRP, and the autoencoder (ENC) portfolio based on the training data (January 1, 2018 – December 31, 2021) are presented in Table 15. Figure 26 depicts the portfolio weights allocated by the portfolios in the form of pie charts. ASIANPAINT received the maximum weights from the MVP and HRP portfolios. The ENC portfolio assigned the highest weight to NTPC.

**Table 15: NIFTY Housing Sector Portfolio Weights Allocation**

| Stock | Portfolio Weights | | |
|---|---|---|---|
| | MVP | HRP | ENC |
| LT | 0.0691 | 0.1287 | 0.1181 |
| ASIANPAINT | 0.3099 | 0.1883 | 0.1149 |
| HDFCBANK | 0.2584 | 0.0882 | 0.0884 |
| ICICIBANK | 0.0005 | 0.0786 | 0.0829 |
| ULTRACEMCO | 0.0594 | 0.0785 | 0.0818 |
| TATASTEEL | 0.0035 | 0.0647 | 0.0709 |
| NTPC | 0.2951 | 0.1754 | 0.1391 |
| HDFC | 0.0009 | 0.0612 | 0.1218 |
| JSWSTEEL | 0.0010 | 0.0465 | 0.0906 |
| GRASIM | 0.0023 | 0.0899 | 0.0915 |

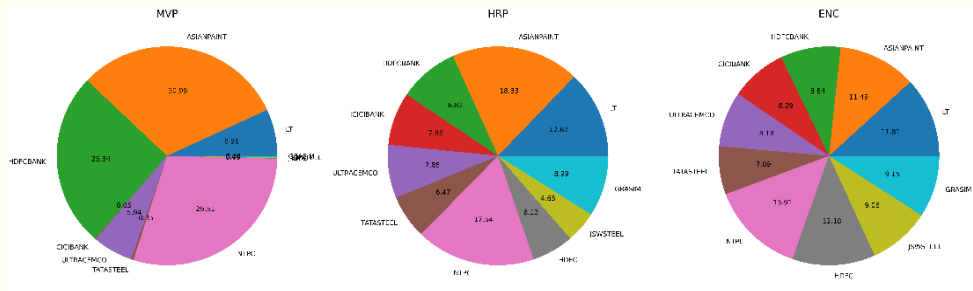

**Figure 26:** Weight allocation to the stocks of the NIFTY housing sector by the MVP, HRP and autoencoder (ENC) portfolios

Figure 27 and Figure 28 show the cumulative returns of the portfolios over the training and the test periods, respectively. In Table 16, the summary of the performances of the three portfolios of the NIFTY housing sector is presented for the training and the test periods, in which the annual returns, annual volatilities (i.e., standard deviations), and the max Sharpe ratio are tabulated.

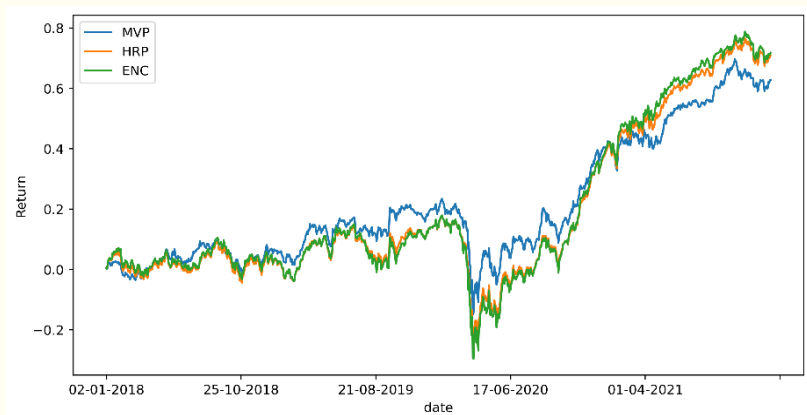

**Figure 27:** Cumulative daily returns yielded by the NIFTY housing sector portfolios for the training period from January 1, 2018, to December 31, 2021

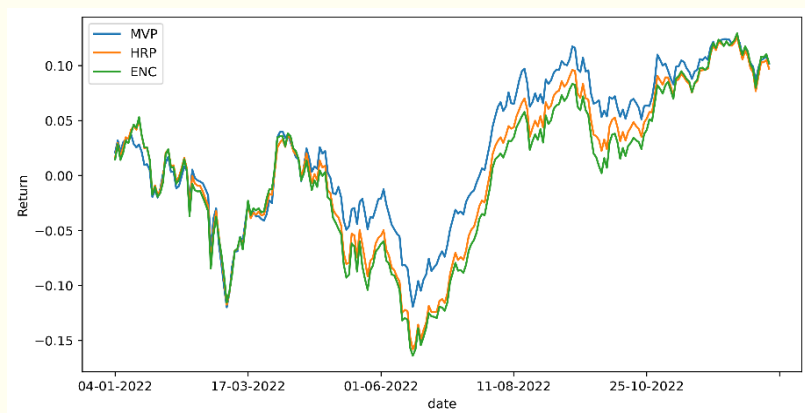

**Figure 28:** Cumulative daily returns yielded by the NIFTY housing sector portfolios for the test period from January 1, 2022, to December 31, 2022

**Table 16: Portfolio Performance on the NIFTY Housing Sector**

| Portfolio | Training Performance | | | Test Performance | | |
|---|---|---|---|---|---|---|
| | Annual Return | Annual Volatility | Sharpe Ratio | Annual Return | Annual Volatility | Sharpe Ratio |
| **MVP** | 16.03% | 19.85 | 0.8079 | 10.38% | 17.61 | 0.5897 |
| **HRP** | 18.09% | 21.74 | 0.8324 | 9.92% | 18.59 | 0.5337 |
| **ENC** | 18.33% | 22.83 | 0.8030 | 10.43% | 19.28 | 0.5411 |

### 5.9 NIFTY Consumption sector

The ten stocks from the NIFTY consumption sector with the maximum free-float market capitalization and their respective contributions to the computation of the overall sectoral index according to the report published by the NSE on Dec 31, 2021, are as follows: ITC (ITC): 11.25, Hindustan Unilever (HINDUNILVR): 10.08, Bharti Airtel (BHARTIARTL): 9.69, Asian Paints (ASIANPAINT): 7.35, Mahindra & Mahindra (M&M): 7.01, Maruti Suzuki (MARUTI): 6.61, Titan Company (TITAN): 5.71, Nestle India (NESTLEIND): 3.84, Britannia Industries (BRITANNIA): 3.04, and Avenue

Supermarts (DMART): 2.81 [3]. The figures mentioned along with the names of the stocks represent the respective weights (in percent) of the stocks used in computing the sectoral index of the NIFTY consumption sector.

The weights assigned by the MVP, HRP, and the autoencoder (ENC) portfolio based on the training data (January 1, 2018 – December 31, 2021) are presented in Table 17. Figure 29 depicts the portfolio weights allocated by the portfolios in the form of pie charts. ITC received the maximum weight from the MVP and HRP portfolios. The ENC portfolio assigned the highest weight to TITAN.

**Table 17: NIFTY Consumption Sector Portfolio Weights Allocation**

| Stock | Portfolio Weights | | |
|---|---|---|---|
| | MVP | HRP | ENC |
| ITC | 0.2412 | 0.1598 | 0.1232 |
| HINDUNILVR | 0.1641 | 0.1232 | 0.1021 |
| BHARTIARTL | 0.0714 | 0.0901 | 0.0958 |
| ASIANPAINT | 0.1027 | 0.1147 | 0.1286 |
| M&M | 0.0299 | 0.0412 | 0.0973 |
| MARUTI | 0.0058 | 0.0446 | 0.0503 |
| TITAN | 0.0477 | 0.0888 | 0.1374 |
| NESTLEIND | 0.1665 | 0.1093 | 0.0904 |
| BRITANNIA | 0.0673 | 0.1223 | 0.0870 |
| DMART | 0.1035 | 0.1061 | 0.0878 |

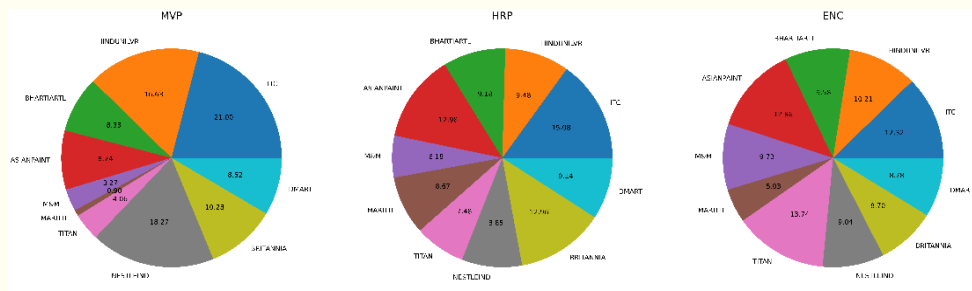

**Figure 29:** Weight allocation to the stocks of the NIFTY consumption sector by the MVP, HRP, and autoencoder (ENC) portfolios

Figure 30 and Figure 31 show the cumulative returns of the portfolios over the training and the test periods, respectively. In Table 18, the summary of the performances of the three portfolios of the NIFTY consumption sector is presented for the training and the test periods, in which the annual returns, annual volatilities (i.e., standard deviations), and the max Sharpe ratio are tabulated.

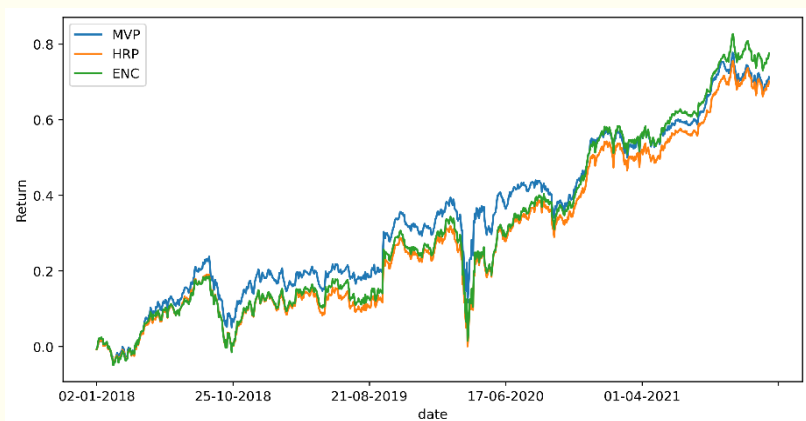

**Figure 30:** Cumulative daily returns yielded by the NIFTY consumption sector portfolios for the training period from January 1, 2018, to December 31, 2021

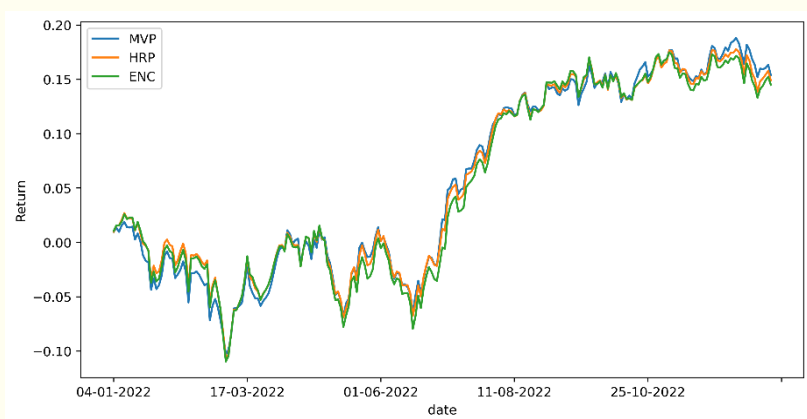

**Figure 31:** Cumulative daily returns yielded by the NIFTY consumption sector portfolios for the test period from January 1, 2022, to December 31, 2022

**Table 18: Portfolio Performance on the NIFTY Consumption Sector**

| Portfolio | Training Performance | | | Test Performance | | |
|---|---|---|---|---|---|---|
| | Annual Return | Annual Volatility | Sharpe Ratio | Annual Return | Annual Volatility | Sharpe Ratio |
| **MVP** | 18.19% | 17.88 | 1.0178 | 15.68% | 15.57 | 1.0074 |
| **HRP** | 17.91% | 18.50 | 0.9680 | 15.18% | 16.13 | 0.9416 |
| **ENC** | 19.81% | 18.74 | 1.0568 | 14.79% | 16.62 | 0.8896 |

### 5.10  NIFTY 100 ESG sector

The ten stocks from the NIFTY 100 ESG sector with the maximum free-float market capitalization and their respective contributions to the computation of the overall sectoral index according to the report published by the NSE on Dec 31, 2021, are as follows: Infosys (INFY): 6.36, Tata Consultancy Services (TCS): 5.99, Housing Development Finance Corporation (HDFC): 4.82, HCL Technologies (HCLTECH): 3.30, ICICI Bank (ICICIBANK): 2.92, Bharti Airtel (BHARTIARTL): 2.81, Tech Mahindra (TECHM): 2.76, Kotak Mahindra Bank (KOTAKBANK): 2.69, Bajaj Finance

(BAJFINANCE): 2.69, and Titan Company (TITAN): 2.69 [3]. The figures mentioned along with the names of the stocks represent the respective weights (in percent) of the stocks used in computing the sectoral index of the NIFTY 100 ESG sector.

The weights assigned by the MVP, HRP, and the autoencoder (ENC) portfolio based on the training data (January 1, 2018 – December 31, 2021) are presented in Table 19. Figure 32 depicts the portfolio weights allocated by the portfolios in the form of pie charts. TCS received the maximum weight from the MVP and HRP portfolios. The ENC portfolio assigned the highest weight to BHARTIARTL.

**Table 19: NIFTY 100 ESG Sector Portfolio Weights Allocation**

| Stock | Portfolio Weights | | |
|---|---|---|---|
| | MVP | HRP | ENC |
| INFY | 0.0895 | 0.1328 | 0.0750 |
| TCS | 0.2571 | 0.1606 | 0.0810 |
| HDFC | 0.0440 | 0.0644 | 0.1180 |
| HCLTECH | 0.0952 | 0.1067 | 0.0862 |
| ICICIBANK | 0.0147 | 0.0485 | 0.1208 |
| BHARTIARTL | 0.1412 | 0.0737 | 0.1209 |
| TECHM | 0.0562 | 0.1131 | 0.1029 |
| KOTAKBANK | 0.1425 | 0.1229 | 0.1175 |
| BAJFINANCE | 0.0044 | 0.0602 | 0.0607 |
| TITAN | 0.1552 | 0.1170 | 0.1169 |

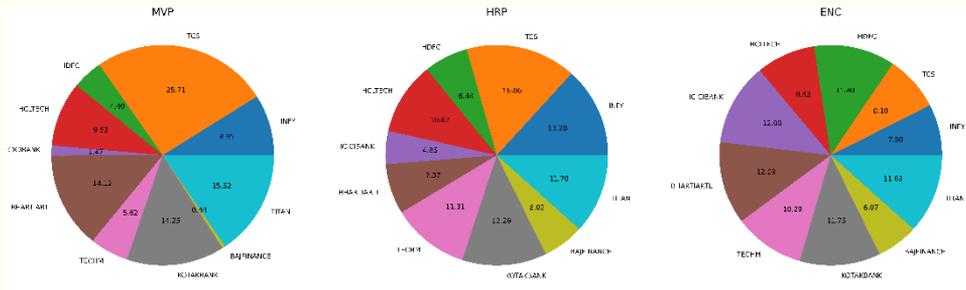

**Figure 32:** Weight allocation to the stocks of the NIFTY 100 ESG sector by the MVP, HRP and autoencoder (ENC) portfolios

Figure 33 and Figure 34 show the cumulative returns of the portfolios over the training and the test periods, respectively. In Table 20, the summary of the performances of the three portfolios of the NIFTY 100 ESG sector is presented for the training and the test periods, in which the annual returns, annual volatilities (i.e., standard deviations), and the max Sharpe ratio are tabulated.

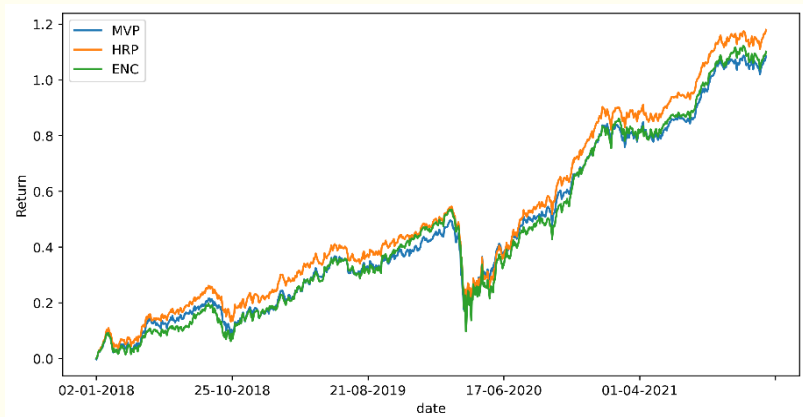

**Figure 33:** Cumulative daily returns yielded by the NIFTY 100 ESG sector portfolios for the training period from January 1, 2018, to December 31, 2021

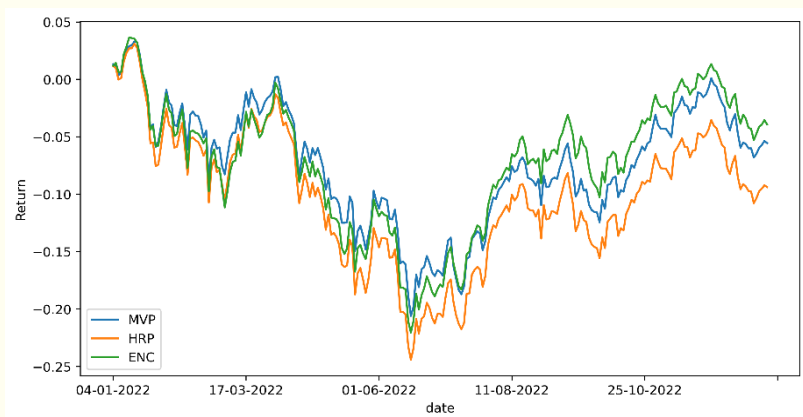

**Figure 34:** Cumulative daily returns yielded by the NIFTY 100 ESG sector portfolios for the test period from January 1, 2022, to December 31, 2022

**Table 20: Portfolio Performance on the NIFTY 100 ESG Sector**

| Portfolio | Training Performance | | | Test Performance | | |
|---|---|---|---|---|---|---|
| | Annual Return | Annual Volatility | Sharpe Ratio | Annual Return | Annual Volatility | Sharpe Ratio |
| **MVP** | 27.73% | 19.68 | 1.4088 | -5.66% | 18.51 | -0.3058 |
| **HRP** | 30.12% | 20.30 | 1.4833 | -9.56% | 19.44 | -0.4921 |
| **ENC** | 28.12% | 21.12 | 1.3292 | -4.00% | 18.84 | -0.2124 |

A summary of the results of the performance of the portfolios on the training data for the ten sectors is exhibited in Table 21. In Table 21, for each sector, the maximum return and Sharpe ratio, and minimum volatility values are presented in bold case. It is observed that the ENC portfolio produced the highest returns for five sectors, while the highest values of the Sharpe ratio were yielded by the MVP portfolio for six sectors. However, the MVP portfolio produced the minimum values of volatility for all ten sectors. Hence, it is concluded that on the training data, the ENC portfolio has exhibited the best performance on the metric annual return, while on the metrics

Sharpe ratio and volatility, the MVP portfolio has performed the best among the three portfolios.

**Table 21: Summary of Portfolio Performance on the Training Data**

| Sector | MVP | | | HRP | | | ENC | | |
|---|---|---|---|---|---|---|---|---|---|
| | Annual Return | Annual Vol | Sharpe Ratio | Annual Return | Annual Vol | Sharpe Ratio | Annual Return | Annual Vol | Sharpe Ratio |
| Commodities | 10.89% | **21.87** | 0.4978 | 13.82% | 23.49 | 0.5883 | **16.59%** | 26.01 | **0.6375** |
| Energy | 19.83% | **20.98** | 0.9451 | 15.81% | 21.92 | 0.7212 | **20.15%** | 23.31 | 0.8644 |
| Manufacturing | 17.40% | **20.76** | 0.8382 | 18.40% | 22.28 | 0.8259 | **19.12%** | 25.58 | 0.7474 |
| Services | 23.89% | **19.81** | **1.2059** | **25.03%** | 21.42 | 1.1684 | 23.65% | 24.31 | 0.9727 |
| MNC | **16.42%** | **18.59** | **0.8833** | 14.73% | 19.49 | 0.7557 | 13.55% | 22.88 | 0.5922 |
| Transportation | **16.26%** | **22.75** | **0.7150** | 12.02% | 23.79 | 0.5053 | 8.55% | 26.03 | 0.3286 |
| Infrastructure | 17.41% | **19.78** | **0.8801** | 17.42% | 20.99 | 0.8300 | 15.70% | 21.11 | 0.7438 |
| Housing | 16.03% | **19.85** | 0.8079 | 18.09% | 21.74 | **0.8324** | **18.33%** | 22.83 | 0.8030 |
| Consumption | 18.19% | 17.88 | 1.0178 | 17.91% | 18.50 | 0.9680 | **19.81%** | 18.74 | **1.0568** |
| ESG | 27.73% | **19.68** | 1.4088 | **30.12%** | 20.30 | **1.4833** | 28.12% | 21.12 | 1.3292 |

The summary of the results of the performance of the portfolios on the test data is exhibited in Table 22, in which, for each sector, the maximum return and Sharpe ratio, and minimum volatility values are presented in bold case. It is observed that the ENC portfolio produced the highest returns for five sectors, while the highest values of the Sharpe ratio were yielded by the MVP portfolio for six sectors. However, the MVP portfolio produced the minimum values of volatility for all eight sectors. Hence, it is concluded that on the test data, the ENC portfolio has exhibited the best performance on the metric annual return, while on the metrics Sharpe ratio and volatility, the MVP portfolio has performed the best among the three portfolios.

**Table 22: Summary of Portfolio Performance on the Test Data**

| Sector | MVP | | | HRP | | | ENC | | |
|---|---|---|---|---|---|---|---|---|---|
| | Annual Return | Annual Vol | Sharpe Ratio | Annual Return | Annual Vol | Sharpe Ratio | Annual Return | Annual Vol | Sharpe Ratio |
| Commodities | **17.51%** | **19.49** | **0.8982** | 13.01% | 20.48 | 0.6354 | 11.74% | 23.32 | 0.5034 |
| Energy | **17.83%** | 19.62 | **0.9086** | 14.82% | **19.54** | 0.7583 | 14.71% | 20.15 | 0.7033 |
| Manufacturing | 9.93% | **16.27** | 0.6102 | 10.99% | 17.61 | **0.6242** | **11.54%** | 20.79 | 0.5549 |
| Services | -0.45% | 18.68 | -0.0240 | 3.40% | 18.53 | 0.1834 | **7.92%** | 19.57 | 0.4047 |
| MNC | **17.18%** | 17.24 | **0.9963** | 15.18% | 17.33 | 0.8754 | 15.70% | 20.04 | 0.7835 |
| Transportation | 25.49% | **21.10** | **1.2081** | 25.73% | 21.51 | 1.1962 | 21.30% | 22.62 | 0.9413 |
| Infrastructure | 10.85% | 17.51 | 0.6200 | 9.70% | 17.57 | 0.5523 | **11.38%** | 17.83 | **0.6384** |
| Housing | 10.38% | **17.61** | 0.5897 | 9.92% | 18.59 | 0.5337 | **10.43%** | 19.28 | 0.5411 |
| Consumption | **15.68%** | **15.57** | **1.0074** | 15.18% | 16.13 | 0.9416 | 14.79% | 16.62 | 0.8896 |
| ESG | -5.66% | **18.51** | -0.3058 | -9.56% | 19.44 | -0.4921 | **-4.00%** | 18.84 | **-0.2124** |

It is interesting to note that the ENC portfolio has yielded the maximum return for the majority of the sectors both on the training and the test data. Hence, for investors looking for higher returns, the best option is to follow the autoencoder-based (ENC) portfolio. However, for those investors who also take into account the risk associated with investments, the MVP portfolio is the best option as this portfolio has yielded the highest risk-adjusted returns for most of the sectors.

It also noted that the transportation sector has produced the highest annual return of 25.73% (which is yielded by the HRP portfolio) among the ten sectors on the test data. The transportation sector has also produced the highest Sharpe ratio (and hence, the highest risk-adjusted return) of 1.2081 (which is yielded by the MVP portfolio) on the test data. The consumption sector exhibited the lowest annual volatility value of 15.57 (which is yielded by the MVP portfolio) among the ten sectors.

# 7. Conclusion

This chapter has presented three portfolio design approaches on ten thematic sectors listed on the NSE of India. The three approaches to portfolio design are the mean-variance portfolio (MVP), hierarchical risk parity (HRP)-based portfolio, and autoencoder-based portfolio. The portfolios are designed based on the historical prices of the ten stocks from the ten sectors which have the maximum free-float market capitalization. The stock prices from January 1, 2018, to December 31, 2021, are used for designing the portfolios. The portfolios are tested on the out-of-sample data of stock prices from January 1, 2022, to December 31, 2022. Three metrics are used in the performance evaluation, annual return, annual risk (i.e., annual volatility), and Sharpe ratios. It is observed that the MVP portfolios have yielded the highest risk-adjusted returns and Sharpe ratios for the majority of the sectors on the out-of-sample data. However, autoencoder-based portfolios are found to have yielded the maximum annual returns for most of the sectors studied in this work. The results indicate that while the autoencoder models are accurate in estimating the future returns of portfolios, these models are inefficient in estimating the future volatilities of the stock prices. In future work, the performance of these portfolios will be studied on stocks listed on the major global stock exchanges.